\documentclass[journal,10pt]{IEEEtran}
\usepackage{amssymb}
\usepackage[cmex10]{amsmath}
\usepackage{stfloats}
\usepackage{graphicx}
\usepackage{subfigure}
\usepackage{tabularx}
\usepackage{epsfig,epsf,color,balance,cite}
\usepackage{verbatim}
\usepackage{url}
\usepackage{bm}

\usepackage{algorithm}
\usepackage{algpseudocode}
\usepackage{multirow}

\begin{document}
\title{A Generalized Data Representation and Training-Performance Analysis for Deep Learning-Based Communications Systems}

\author{
\IEEEauthorblockN{
        Xiao Chen,
        ~Liang Wu, \IEEEmembership{Member, IEEE},
        and ~Zaichen Zhang, \IEEEmembership{Senior Member, IEEE}
        }
\thanks{X. Chen, L. Wu and Z. Zhang are with National Mobile Communications Research Laboratory, Southeast University, Nanjing,
210096, China (email: \{chen$\_$xiao, wuliang, zczhang\}@seu.edu.cn). Zaichen Zhang is the corresponding author.}
}

\maketitle

\begin{abstract}
Deep learning (DL)-based autoencoder is a potential architecture to implement end-to-end communication systems.
In this letter, we first give a brief introduction to the autoencoder-represented communication system.
Then, we propose a novel generalized data representation (GDR) aiming to improve the data rate of DL-based communication systems.
Finally, simulation results show that the proposed GDR scheme has lower training complexity, comparable block error rate performance and higher channel capacity than the conventional one-hot vector scheme.
Furthermore, we investigate the effect of signal-to-noise ratio (SNR) in DL-based communication systems and prove that training at a high SNR could produce a good training performance for autoencoder.
\end{abstract}

\begin{IEEEkeywords}
Autoencoder, communication systems, data representation, deep learning, SNR.
\end{IEEEkeywords}

\section{Introduction}
To ensure dramatical demand for various applications and services, the next-generation network must be able to deliver enhanced mobile broadband, ultra-reliable and low-latency communications (URLLC), and massive Internet of Things (IoT) ecosystems \cite{6736746, 7414384, 7041045, 7736615}.
The primary concern is to satisfy the exponential rise in the number of user equipments and the traffic capacity of future communication systems.
Hence, several promising technologies have been proposed including massive multi-input and multi-output (MIMO) transmissions, millimeter wave (mmWave) communications, ultra-dense networks (UDNs), etc.
However, there exist a number of limitations for conventional communication systems including unavailable channel estimation of complex scenario, high time complexity to processing big data, and sub-optimal performance caused by conventional block-structure \cite{8214233,8054694,01151, 8233654}.
As a promising technique, deep learning (DL) applies a useful and insightful way to implement communication systems using neural networks (NNs).
Different from conventional communication system, DL-based communication system is a novel concept going back to the original definition of the communication system.
Its goal is to optimize transmitter and receiver jointly for end-to-end performance without block-structure.
Furthermore, DL-based communication systems can be adapted to a practical system over any type of channel, hold a promise for efficiency due to the highly parallelized architectures implementing the computation offline, and so on \cite{8054694, 8214233}.
Attracted by these advantages, several research groups \cite{07738, 08044, 8242643, 8322184} have recently begun to investigate the DL-based communications and signal processing using state-of-the-art tools and hardware.

However, in the previous studies \cite{8054694, 8233654, 07738, 08044}, one-hot vector as the main data representation has a low data rate in conventional DL-based communication systems.
For example, an $M\times 1$ one-hot vector consists of 0s in all cells with the exception of a single 1 in a cell \cite{176685}, such as, $[0,\cdots,0,1,0,\cdots,0]^T$, which results in limited transmitted bits since there are only $C_M^1$ transmitted messages.
On the other hand, the DL-based communication system is represented and implemented by autoencoder which is trained over dataset offline. Then, the trained autoencoder will be applied for practical systems.
In general, the autoencoder is trained under a fixed signal-to-noise ratio (SNR), while it is expected for working well in a wide SNR region.
It is mentioned that different trained SNR will generate various training performance for autoencoder \cite{8054694}, while, there is no study on the effect of SNR until now.
Therefore, our objective is not only to design a new data representation to improve transmission rate, but also to investigate the effect of trained SNR aiming to improve the system performance of DL-based communication autoencoder.

In this letter, we propose a generalized data representation (GDR) scheme to improve the data rate and analyze the effect of SNR in DL-based communication systems.
Then, simulation results demonstrate low complexity, good block error rate (BLER) performance and high channel capacity achieved by the proposed GDR scheme.
To the best of our knowledge, the GDR scheme is proposed and verified its effectiveness for the first time in this letter.
Further, it is analyzed and proved that a high trained SNR can provide an excellent training performance for the autoencoder, which provides a reliable reference for selecting the trained SNR.

\section{Deep Learning-Based Communication Systems}
In this section, we describe the DL-based autoencoder for end-to-end communication systems and then provide the problem formulation of our research issues.
\vspace{-1em}
\subsection{Autoencoder for End-to-End Communication Systems}
\begin{figure*}[t]
  \centering
  \epsfig{file=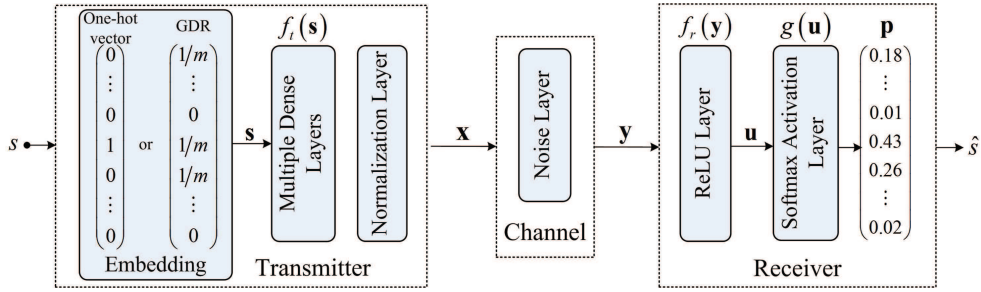,width=13cm}
  \vspace{-1em}
  \caption{A DL-based communication system represented as an autoencoder \cite{8054694}.}
  \label{Sys}
\end{figure*}
We consider a DL-based communication system represented as an autoencoder\footnote{The autoencoder describes a deep NN that applies unsupervised learning aiming to reconstruct input at the output \cite[Ch. 14]{Goodfellow-et-al-2016}.} consisting of transmitter, channel, and receiver, as shown in Fig. \ref{Sys}.
At the transmitter, the message $s\in \{1, 2, \cdots, M\}$ is first transformed to a vector $\mathbf s\in \mathbb{R}^{M}$ after the embedding processing.
For example, while the message $s=2$ is transmitted, the corresponding output of embedding is a one-hot vector $\mathbf s=[0,1,0,\cdots,0]^T$ in conventional DL-based communication systems.
Then, the multiple dense layers including linear layer and rectified linear unit (ReLU) layer apply the transformation $f_t: \mathbb{R}^{M}\mapsto\mathbb{R}^{n}$ to produce the transmitted signal for $n$-discrete channel uses \cite{8214233}.
Finally, the normalization layer ensures the power constrain of the transmitted signal as $\mathbb{E}\{|x_j|^2\}\leq1$ ($j=1,\cdots,n$).

The channel is implemented by a noise layer with its output being the received signal $\mathbf y$ that can be given by
\begin{IEEEeqnarray}{rCl}
\label{y}
\mathbf{y}=\mathbf x+\mathbf n\textrm{,}
\end{IEEEeqnarray}
where $\mathbf{n}\sim\mathcal{CN}(\mathbf{0},\sigma^2\mathbf{I}_n)$ denotes additive white Gaussian noise (AWGN) vector with a fixed variance $\sigma^2=(2RE_b/N_0)^{-1}$, $R$ is the data rate, $E_b$ is the energy per bit, and $N_0$ denotes the noise power spectral density.
Notably, there is no complex operation in existing NN architectures, thus, the complex number is represented by two real number \cite{8054694}.
Here, we assume that channel coefficients are real-valued.

At the receiver, the received signal $\mathbf y$ is passing through the ReLU layer to realize the transformation $f_r: \mathbb{R}^{n}\mapsto\mathbb{R}^{M}$.
The last layer of the receiver has a softmax activation which is a generalization of the logistic function that compresses an $M$-dimensional vector of arbitrary real values to an $M$-dimensional probability vector $\mathbf p$, where each element is in the range (0, 1], and all the elements add up to 1 \cite{Goodfellow-et-al-2016}.
In the conventional autoencoder scheme with one-hot vector, the estimated message $\hat{s}$ is derived from the index of element with the highest probability in $\mathbf p$.
The autoencoder of communication system can be trained on a large training dataset according to the categorical cross-entropy loss function.

\vspace{-1em}
\subsection{Problem Formulation}
One-hot vector is the conventional data representation with only one non-zero element.
Thus, the data rate of the conventional DL-based communication system is limited to be
\begin{IEEEeqnarray}{rCl}
\label{Roh}
R_C=\frac{\left\lfloor\log_2M\right\rfloor}{n} \quad(\textrm{bits/channel use})\textrm{.}
\end{IEEEeqnarray}
Over the last few years, the demand for high data rates has experienced an unprecedented growth in communication systems \cite{7414384}.
Therefore, providing a high data rate is a crucial issue for DL-based communication systems.
From the definition of the data rate $R_{\emph{def}}=\frac{\textrm{Number of bits}}{\textrm{Channel uses}}$, it is obvious that, for the same channel environment, the data rate is proportional to the number of bits which are conveyed.
So, a new data transmission scheme is required to meet the high data rate demands in communication systems.

On the other hand, the autoencoder of communication system is trained over a fixed SNR offline, while it is expected for a great generalization performance that can be used for wide SNR region.
An improper trained SNR could result in the performance degradation of DL-based communication systems \cite{8054694}.
In consequence, the effect of the SNR need be investigated to provide a reliable criterion for selecting a proper SNR while training the autoencoder.

\section{Proposed Data Representation and Performance Analysis}
In this section, we propose a generalized data representation scheme and investigate the effect of SNR for DL-based communication systems.
\vspace{-1em}
\subsection{Generalized Data Representation}
Instead of the conventional one-hot vector, bit vector is firstly considered to improve the data rate for DL-based communication systems.
An $m$-order bit vector $\mathbf b\in \mathbb{R}^M$ is defined as
\begin{IEEEeqnarray}{rCl}
\label{b}
\mathbf{b}={\underbrace {[1\quad 0\quad \cdots\quad 0\quad 1\quad \cdots\quad 1\quad 0]}_{\textrm{$m$ of $M$ are 1}}}^T\textrm{,}
\end{IEEEeqnarray}
where $m=1,2,\cdots, \left\lfloor M/2\right\rfloor$ and $\lfloor\cdot\rfloor$ denotes the floor operation.
The bit vector provides $C_M^m$ possible messages for the transmission.
In general, the number of possible symbols in the constellation diagram is a power of 2.
Therefore, we only select $2^{\left\lfloor\log_2C_M^m\right\rfloor}$ out of $C_M^m$ possible symbols for the communications.

Furthermore, in information theory, cross-entropy is defined by two probability distributions over the same underlying set of events \cite{Kullback59}.
For the autoencoder shown in Fig. \ref{Sys}, the categorical cross-entropy loss function is given by
\begin{IEEEeqnarray}{rCl}
\label{L}
L(\mathbf s,\mathbf p)=-\sum_is_i\log(p_i)\textrm{,}
\end{IEEEeqnarray}
where $i=1, 2, \cdots, M$ is the element index of vector, the output of embedding $\mathbf s$ is the practical probability distribution, and the probability vector $\mathbf p$ is the estimated probability distribution to be optimized.
The training goal of the autoencoder is minimizing the categorical cross-entropy loss function.

Thus, combining the form of bit vector, we propose a generalized data representation as a probability distribution
\begin{IEEEeqnarray}{rCl}
\label{s}
\mathbf{s}={\underbrace {\left[\frac{1}{m} \quad 0\quad \cdots\quad 0\quad \frac{1}{m} \quad \cdots\quad \frac{1}{m} \quad 0\right]}_{\textrm{$m$ of $M$ are }\frac{1}{m}}}^T\textrm{,}
\end{IEEEeqnarray}
where the estimated message $\hat{s}$ can be derived from the indexes of elements with the $m$ highest probabilities in $\mathbf p$.
The conventional one-hot vector is a special case of the proposed GDR scheme when $m=1$.

With this proposed GDR, first, the data rate of DL-based communication systems can be improved to
\begin{IEEEeqnarray}{rCl}
\label{R}
{R}=\frac{\left\lfloor\log_2C_M^m\right\rfloor}{n}\quad(\textrm{bits/channel use})\textrm{.}
\end{IEEEeqnarray}
While $m=1$, the data rate of conventional one-hot vector in (\ref{Roh}) is obtained.
The data rate increases with $m$, while $M$ keeps constant.
The performance gain of the proposed GDR scheme will greatly increase when the vector size $M$ increases.

Second, the channel capacity of the proposed GDR scheme in the DL-based communication system is derived as
\begin{IEEEeqnarray}{rCl}
\label{C}
{C}&&=\log_2(1+\textrm{SNR})=\log_2(1+\frac{1}{\sigma^2})\\
&&=\log_2\left(1+\frac{2E_b\cdot\left\lfloor\log_2C_M^m\right\rfloor}{N_0\cdot n}\right)\quad(\textrm{bits/s/Hz})\nonumber \textrm{.}
\end{IEEEeqnarray}
It is obvious that the capacity can be improved by using the proposed GDR in DL-based communication systems.

\vspace{-1em}
\subsection{The Effect of SNR to System Performance}
With the noise layer, the effect of SNR needs to be considered in DL-based communication systems.
In Fig. \ref{Sys}, the output of the ReLU layer at receiver can be written as
\begin{IEEEeqnarray}{rCl}
\label{u}
\mathbf u=\mathbf A (\mathbf x+\mathbf n)\textrm{,}
\end{IEEEeqnarray}
where $\mathbf A\in \mathbb{R}^{M\times n}$ realizes the transformation $f_r: \mathbb{R}^{n}\mapsto\mathbb{R}^{M}$.
From $\mathbf{n}\sim\mathcal{CN}(\mathbf{0},\sigma^2\mathbf{I}_n)$, it can be derived that $\mathbf{u}\sim\mathcal{CN}(\mathbf {Ax} ,\sigma^2\mathbf{AA}^T)$.
Thus, the $i$th element $u_i$ is a normal distribution with its expectation and variance given by
\begin{IEEEeqnarray}{rCl}
\label{ED}
\left\{
\begin{array}{ll}
\textrm{E}(u_i)=\mathbf a_i\mathbf x\\
\textrm{D}(u_i)=\sigma^2\mathbf a_i\mathbf a_i^T\\
\end{array}\textrm{,}
\right.
\end{IEEEeqnarray}
respectively, where $\mathbf a_i\in \mathbb{R}^{1\times n}$ denotes the $i$th row of $\mathbf A$.

Next, a probability vector is derived from the softmax function with its $i$th element as
\begin{IEEEeqnarray}{rCl}
\label{softmax}
p_i=\frac{e^{u_i}}{\sum_k^{M}e^{u_k}}\textrm{.}
\end{IEEEeqnarray}
It is clear that the probability mainly depends on numerator with a normalizer at denominator.
From (\ref{ED}), $e^{u_i}$ is a log-normal distribution with the expectation and variance being
\begin{IEEEeqnarray}{rCl}
\label{logED}
\left\{
\begin{array}{ll}
\textrm{E}(e^{u_i})=e^{\mathbf a_i\mathbf x+\frac{\sigma^2\mathbf a_i\mathbf a_i^T}{2}}\\
\textrm{D}(e^{u_i})=(e^{\sigma^2\mathbf a_i\mathbf a_i^T}-1)e^{2\mathbf a_i\mathbf x+\sigma^2\mathbf a_i\mathbf a_i^T}\\
\end{array}\textrm{,}
\right.
\end{IEEEeqnarray}
respectively.
When the $\mathrm{SNR}=1/\sigma^2$ decreases (means the noise variance $\sigma^2$ increases), the variance of $e^{u_i}$ increases according to (\ref{logED}).
The increased variance indicates that $e^{u_i}$ is spreading out from the expected value, which will lead to a worse estimated probability vector $\mathbf{p}$.
It can be seen that, $p_i$, the probability of the $i$th element is directly affected by the SNR.
The effect of trained SNR will also be analyzed through simulations.

\vspace{-1.3em}
\section{ Numerical Results}
\begin{table}[t]
\caption{Parameters for the autoencoder setup}
\vspace{-1.5em}
\begin{center}
\begin{tabular}{l|l}
\hline
\textbf{Parameter} & \textbf{Value} \\
\cline{1-2}
Optimizer&  Adam \cite{adam}\\
Loss function & Categorical cross-entropy \\
Epoch &  150 \\
Batch size & 45\\
Trained samples &2$\times10^4$\\
Test samples &1$\times 10^6$\\
\hline
\end{tabular}
\label{tab1}
\end{center}
\end{table}

In this section, we evaluate the performance of the proposed GDR scheme in the DL-based communication system via simulations on the TensorFlow framework.
In all simulations, the autoencoder is trained over the stochastic AWGN channel model with $n=7$ channel uses without exhaustive hyperparameter tuning.
Here, we use the same set of parameters for the autoencoder setup as described in TABLE \ref{tab1}.

\newcommand{\tabincell}[2]{\begin{tabular}{@{}#1@{}}#2\end{tabular}}
\begin{table}[t]
\caption{Training parameters of autoencoder}
\vspace{-1.5em}
\begin{center}
\begin{tabular}{c|c|c|c|c|c}
\hline
\tabincell{c}{Trainable \\parameters} &\tabincell{c}{Multiple \\dense layers} &\tabincell{c}{Normaliz-\\ation layer} &\tabincell{c}{ReLU \\layer} &\tabincell{c}{Softmax \\layer} &Total\\
\hline
M=8      &135  &14  &64  &72   &285\\
\hline
M=16    &391   &14  &128  &272  & 805\\
\hline
M=64    &4615   &14  &512 &4160  &9301\\
\hline
\end{tabular}
\label{tab2}
\end{center}
\end{table}

TABLE \ref{tab2} presents the number of practical parameters in autoencoder training, where different size of the data representation is employed.
It is clear that the number of trainable parameters increases with $M$ from 8 to 64, not only for the total number but also for the number of each layer, which leads to an increasing complexity for training.
For the conventional one-hot vector, the data rate can be improved by increasing $M$ at the cost of high complexity.
While, the data rate of the proposed GDR scheme can be improved by controlling the number of non-zero elements $m$ with a small $M$ as shown in (\ref{R}), which leads to a low complexity.

\begin{figure}[t]
\begin{center}
\epsfig{file=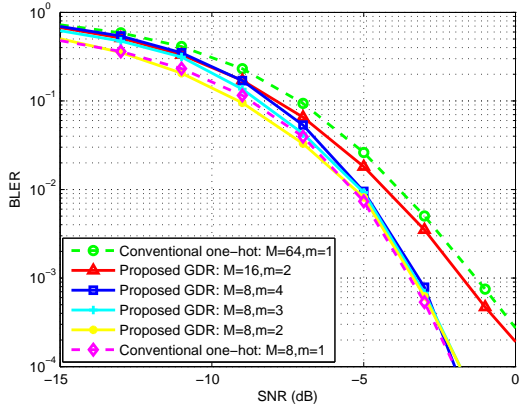,width=7.8cm}
\vspace{-1em}
\caption{Simulated BLER for the autoencoder with different data representations, while the trained SNR is 0 dB.}
\label{BLER1}
\end{center}
\end{figure}

Fig. \ref{BLER1} shows the simulated BLER performance of DL-based communication system employing proposed GDR and conventional one-hot vector schemes for comparison, while the trained SNR is 0 dB.
1) In Fig. \ref{BLER1}, with the same data rate $R=6/7$ (bits/channel use) including $M=8$ with $m=4$, $M=16$ with $m=2$, and $M=64$ with $m=1$, the BLER of the proposed GDR scheme is better than that of conventional one-hot vector.
Obviously, the BLER decreases with the vector size $M$, since the smaller $M$ requests less trainable parameters as in TABLE \ref{tab2}.
2) From Fig. \ref{BLER1}, with the same vector size M=8, the proposed GDR schemes ($m=4$, $m=3$, and $m=2$) obtain a comparable BLER performance compared with the conventional one-hot vector scheme ($M=8$, $m=1$).
3) The simulated BLER is less than $10^{-3}$ when the SNR is greater than 0 dB in Fig. \ref{BLER1}, which demonstrates that the autoencoder obtained a high accuracy by sufficient training.

\begin{figure}[t]
\begin{center}
\epsfig{file=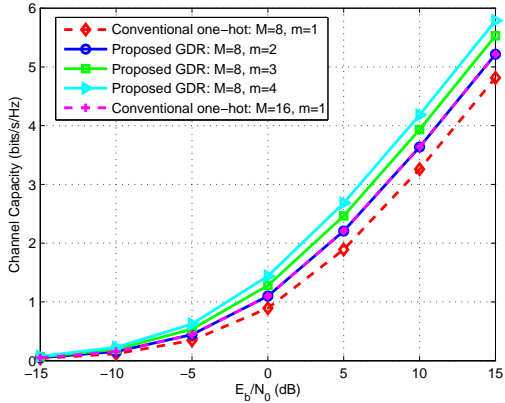,width=7.8cm}
\vspace{-1em}
\caption{Channel capacity for the autoencoder with different data representations, while the trained SNR is 0 dB.}
\label{MSP}
\end{center}
\end{figure}

Fig. \ref{MSP} illustrates the channel capacity of the DL-based communication system employing different data representations, while the trained SNR is 0 dB.
It can be seen from Fig. \ref{MSP} that, with $M=8$, the channel capacity is increased when the order $m$ increases from 1 to 4.
It is proved that the proposed GDR scheme can obtain a remarkable channel capacity improvement.
Besides, the channel capacity of proposed GDR scheme ($M=8$, $m=2$) is the same as that of the conventional one-hot vector scheme ($M=16$, $m=1$) in Fig. \ref{MSP}.
To obtain the same capacity with GDR scheme, the conventional one-hot vector scheme need to improve the vector size $M$, which will lead to a BLER performance degradation due to the high complexity.

\begin{figure}[t]
\begin{center}
\epsfig{file=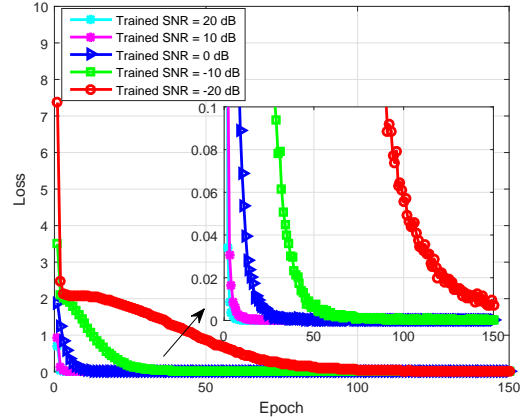,width=7.8cm}
\vspace{-1em}
\caption{Simulated loss function performance of autoencoder while training at different SNRs.}
\label{TSNR}
\end{center}
\end{figure}

Fig. \ref{TSNR} depicts the simulated loss function performance of DL-based communication system while training at different SNRs.
Hear, an epoch is the process that training dataset totally passes through the autoencoder once.
In Fig. \ref{TSNR}, with the trained SNR increasing form -20 dB to 20 dB, the loss value decreases and the convergence of loss function gets better, which indicates that the good channel environment contributes to improve the training performance.
The simulated results prove the corresponding analysis in Section III-B.

\vspace{-0.5em}
\section{Conclusion}
In this letter, we have proposed a generalized data representation scheme to address the problem of limited data rate for DL-based communication systems.
Simulation results show that the proposed GDR scheme can obtain low complexity and good BLER performance, as well as effectively improve the channel capacity compared with the conventional one-hot vector scheme.
Besides, it is analyzed and verified that SNR will affect the training performance for autoencoder greatly.

\bibliographystyle{IEEEtran}
\bibliography{IEEEabrv,DL}

\end{document}